\documentclass[aps,preprint,nofootinbib,superscriptaddress]{revtex4}
\usepackage{amssymb}

\usepackage{epsfig}

\begin{document}

\date{\today}
\title{Applications of Skyrme energy-density functional to fusion reactions spanning the fusion barriers}

\author{Min Liu}
\affiliation{China Institute of Atomic Energy, Beijing 102413, P.
R. China}

\author{Ning Wang}
\email{Ning.Wang@theo.physik.uni-giessen.de}
\affiliation{Institute of Theoretical Physics, Chinese Academic of
Science, Beijing 100080, P. R. China}

\author{Zhuxia Li}
\email{lizwux@iris.ciae.ac.cn} \affiliation{China Institute of
Atomic Energy, Beijing 102413, P. R. China} \affiliation{Institute
of Theoretical Physics, Chinese Academic of Science, Beijing
100080, P. R. China} \affiliation{Nuclear Theory Center of
National Laboratory of Heavy Ion Accelerator, Lanzhou 730000, P.
R. China}

\author{Xizhen Wu}
\affiliation{China Institute of Atomic Energy, Beijing 102413, P.
R. China} \affiliation{Nuclear Theory Center of National
Laboratory of Heavy Ion Accelerator, Lanzhou 730000, P. R. China}

\author{Enguang Zhao}
\affiliation{Institute of Theoretical Physics, Chinese Academic of
Science, Beijing 100080, P. R. China} \affiliation{Department of
Physics, Tsinghua University, Beijing 100084, P. R. China}

\begin{abstract}
The Skyrme energy density functional has been applied to the study
of heavy-ion fusion reactions. The barriers for fusion reactions
are calculated by the Skyrme energy density functional with proton
and neutron density distributions determined by using restricted
density variational (RDV) method within the same energy density
functional together with semi-classical approach known as the
extended semi-classical Thomas-Fermi method. Based on the fusion
barrier obtained, we propose a parametrization of the empirical
barrier distribution to take into account the multi-dimensional
character of real barrier and then apply it to calculate the
fusion excitation functions in terms of barrier penetration
concept.  A large number of measured fusion excitation functions
spanning the fusion barriers can be reproduced well. The
competition between suppression and enhancement effects on
sub-barrier fusion caused by neutron-shell-closure and excess
neutron effects is studied.

\end{abstract}

\maketitle


\begin{center}
\textbf{I. INTRODUCTION}
\end{center}
A large number of fusion excitation functions have been
accumulated in recent decades\cite{3,13,20}, which provides a
possibility for a systematic study on fusion reactions. Newton
et.al. \cite{New04} analyzed a total of 46 fusion excitation
functions at energies above the  average fusion barriers using the
Woods-Saxon form for the nuclear potential in a barrier passing
model of fusion. They found that the empirical diffuseness
parameters $a$ ranging between 0.75 and 1.5 were considerably
larger than those obtained from  elastic scattering data and the
deduced $a$ showed  strong increase with increasing the charge
product $Z_1Z_2$. Thus, it results in a certain difficulty for
giving satisfied predictions of fusion cross sections for
unmeasured reaction systems. The fusion coupled channel model is
successful for describing fusion excitation function of heavy-ion
reaction at energies near fusion barrier. However, with the
increasing of the neutron excess and the charges (the product
$Z_1Z_2$) of two nuclei, the fusion coupled channel model
encounters a lot of difficulties due to a very large number of
degrees of freedom involved. In order to carry out a systematic
study of fusion excitation functions, a simple and useful approach
seems to be required. A semi-empirical approach in which the
quantum penetrability of Coulumb barrier is calculated by using
the concept of barrier distribution arising due to the
multi-dimensional character of the real nucleus-nucleus
interaction\cite{Zag02,Zag03,Stel88} is very helpful in this
aspect. Thus, a reasonable parametrization of the weighing
function describing the barrier distribution based on the
interaction potential in the entrance channel seems to be very
useful.

In \cite{Deni02} the interaction potential in the entrance channel
for fusion reactions was calculated based on the semi-classical
expressions of the Skyrme energy density functional
\cite{Vau72,brack,Bart02}. In the calculation of the interaction
potential, which is the difference between the energy of the total
system and the energies of individual projectile and target, the
density distributions of the projectile and target are required.
In \cite{Deni02} the density distributions of the projectile and
target were determined by Hartree-Fock-Bogoliubov calculations.
According to the Hohenberg and Kohn theorem\cite{Hoh64}, the
energy of a N-body system of interaction fermions is a unique
functional of local density.  In the framework of the
semi-classical Extended Thomas Fermi (ETF) approach together with
a Skyrme effective nuclear interaction such a functional can be
derived systematically. The density functional theory is widely
used in the study of the nuclear ground state which provides us
with a useful balance between accuracy and computation cost
allowing large systems with a simple self-consistent manner. Thus,
it is very suitable to perform the calculation of the
entrance-channel potential and the densities of reaction partners
based on the same Skyrme energy density functional. Of course, the
different density distributions obtained by different approaches
will influence the entrance-channel potential. While a
self-consistent treatment for both density distribution and the
entrance-channel potential seems to be more reasonable.

In this work, the Skyrme energy density functional is applied to
make a systematic study of fusion reactions. Firstly, we will use
the semi-classical expressions of the Skyrme energy density
functional to study the energies and the density distributions of
a series of nuclei by the restricted density variational (RDV)
method\cite{Bart85,brack,Cen90,Bart02}. Secondly, with the density
distributions obtained, the entrance-channel potentials of a
series of fusion reactions are calculated. Then, based on the
entrance-channel potential obtained, a parametrization of the
empirical barrier distribution is proposed to take into account
the multi-dimensional character of real barrier and then apply it
to calculate the fusion excitation functions in terms of barrier
penetration concept. The paper is organized as follows: In Sec.II,
the properties of ground state of nuclei and the entrance-channel
potential are studied in the framework of the Skyrme energy
density functional. In Sec.III, an approach to calculate fusion
excitation functions is introduced and a large number of
calculated results are presented. Finally, the summary and
discussion are given in Sec.IV.

\begin{center}
\textbf{II. THE CALCULATIONS OF FUSION BARRIER}
\end{center}
The energy density functional theory is widely used in many-body
problems. In the framework of the semi-classical Extended Thomas
Fermi (ETF) approach together with a Skyrme effective nuclear
interaction, the energy density functional can be derived
systematically. We take the Skyrme Hartree-Fock (Skyrme HF)
formalism of the energy density functional\cite{Vau72,brack} and
based on it we calculate the proton and neutron densities of
nuclei by means of restricted density variational
method\cite{Bart85,brack,Cen90,Bart02}. With the neutron and
proton densities determined in this way we calculate the fusion
barrier for fusion reaction based on the same energy density
functional.

\begin{center}
\textbf{A.  Skyrme Energy Density Functional}
\end{center}
The total binding energy of a nucleus can be expressed as the
integral of energy density functional \cite{Vau72,Bart02}
\begin{eqnarray}
E = \int {\mathcal H} \; d{\bf r}.
\end{eqnarray}
The energy density functional $\mathcal H$ includes the kinetic,
nuclear interaction and Coulomb interaction energy parts
\begin{equation}
{\mathcal H} = \frac{\hbar^2}{2m} [\tau_p({\bf r}) +\tau_n({\bf
r})] + {\mathcal H}_{\rm sky}({\bf r}) + {\mathcal H}_{\rm
coul}({\bf r}).
\end{equation}
For the kinetic energy part, the extended Thomas-Fermi (ETF)
approach including all terms up to second order in the spatial
derivatives (ETF2), is applied as that was done in
ref.\cite{Deni02}. With the effective-mass form factor
\cite{Bon87}
\begin{equation}
f_{i}({\bf r}) =1+\frac{2m}{\hbar^2} \left\{
 \frac{1}{4}\left[t_1(1+\frac{x_1}{2})+t_2(1+\frac{x_2}{2})\right]
\rho({\bf r})+
\frac{1}{4}\left[t_2(x_2+\frac{1}{2})-t_1(x_1+\frac{1}{2}) \right]
\rho_i ({\bf r}) \right\},
\end{equation}
the kinetic energy densities $\tau$ for protons ($i=p$) and
neutrons ($i=n$) are given by
\begin{eqnarray} \tau_{i}({\bf r}) = \frac{3}{5}(3\pi^2)^{2/3} \rho_{i}^{5/3} + \frac{1}{36} \frac{(
\nabla \rho_{i} )^2}{ \rho_{i} } + \frac{1}{3} \Delta \rho_{i}
+\frac{1}{6} \frac{ \nabla \rho_{i} \nabla f_{i} + \rho_{i} \Delta
f_{i} }{ f_{i} } \\
- \frac{1}{12} \rho_{i} \left( \frac{\nabla f_{i}}{f_{i}}
\right)^2 \nonumber + \frac{1}{2} \rho_{i} \; \left(
\frac{2m}{\hbar^2} \; \frac{W_0}{2} \; \frac{ \nabla (\rho +
\rho_{i}) }{f_{i}} \right)^2 \nonumber ,
\end{eqnarray}
where $\rho_{i}$ denotes the proton or neutron density of the
nucleus and $\rho=\rho_p+\rho_n$, $W_0$ denotes the strength of
the Skyrme spin-orbit interaction.  The nuclear interaction part
with Skyrme interaction ${\mathcal H}_{\rm sky}$ reads
\begin{eqnarray}
{\mathcal H}_{\rm sky}({\bf r}) \; = \; \frac{t_0}{2} \;
[(1+\frac{1}{2}x_0) \rho^2 -
(x_0+\frac{1}{2}) (\rho_p^2+\rho_n^2)] \;\\
+\frac{1}{12} t_3 \rho^\alpha [(1+\frac{1}{2}x_3 )\rho^2 -
(x_3+\frac{1}{2}) (\rho_p^2+\rho_n^2) ] \nonumber \\
+\frac{1}{4} [t_1(1+\frac{1}{2}x_1)+t_2(1+\frac{1}{2}x_2)] \tau
\rho
\nonumber \\
+\frac{1}{4} [t_2(x_2+\frac{1}{2}) - t_1(x_1+\frac{1}{2})] (\tau_p
\rho_p+\tau_n \rho_n)
\nonumber \\
+\frac{1}{16}[3t_1(1+\frac{1}{2} x_1)-t_2(1+\frac{1}{2}x_2)]
(\nabla
\rho)^2 \nonumber \\
- \frac{1}{16}[3t_1(x_1+\frac{1}{2} )+t_2(x_2+\frac{1}{2})]
[(\nabla
\rho_n)^2 +(\nabla \rho_p)^2 ] \nonumber \\
- \frac{W_0^2}{4} \frac{2m}{\hbar^2}
\left[\frac{\rho_p}{f_p}(2\nabla \rho_p+\nabla \rho_n)^2+
\frac{\rho_n}{f_n}(2\nabla \rho_n+\nabla \rho_p)^2 \right] ,
\nonumber
\end{eqnarray}
where $t_0$, $t_1$, $t_2$, $t_3$, $x_0$, $x_1$, $x_2$, $x_3$,and
$\alpha$ are Skyrme-force parameters\cite{Deni02,Bart02}. The last
term in the right hand of expression (5) is the semi-classical
expansions (up to second order in $\hbar$) of spin-orbit
densities\cite{Vau72}. The Coulomb energy density can be written
as the sum of the direct and exchange contribution, the latter
being taken into account in the Slater approximation,
\begin{eqnarray}
{\mathcal H}_{\rm Coul}({\bf r}) = \frac{e^2}{2} \rho_p ({\bf r})
\int \; \frac{\rho_p ({\bf r}' )}{|{\bf r}-{\bf r}' |} d {\bf r}'
-\frac{3e^2}{4} \left( \frac{3}{\pi} \right)^{1/3} (\rho_p({\bf
r}))^{4/3}.
\end{eqnarray}
From the above expressions (1)--(6), one can see that the total
energy of a nuclear system can be expressed as a functional of
protons and neutrons densities [$\rho_p({\bf r})$, $\rho_n({\bf
r})$] under the Skyrme interaction associated with the ETF
approximation.

\begin{center}
\textbf{B. The Neutron and Proton Densities of Nuclei}
\end{center}
By minimizing the total energy of the system given by expression
(1), the neutron and proton densities can be obtained, that is to
solve the variational equation
\begin{equation}
\frac{\delta}{\delta \rho_{i}} \int \left\{ {\mathcal
H}[\rho_n({\bf r}), \rho_p({\bf r})] - \lambda_n \rho_{n}({\bf r})
    - \lambda_{p} \rho_{p}({\bf r}) \right\} d{\bf r}=0,
\end{equation}
with the Lagrange multipliers $\lambda_n$ and $\lambda_p$ to
ensure the conservation of neutron and proton number. This density
variational problem has been solved in two different ways in the
past: either by resolving the Euler-Lagrange equation
\cite{Bart85,Cen90} resulting from Eq.(7) or by carrying out the
variational calculation in a restricted subspace of functions
\cite{brack,Bart85,Cen90,Bart02}. In this work we take the neutron
and proton density distributions of a nucleus as spherical
symmetric Fermi functions
\begin{equation}
\rho_{i}({\bf r})=\rho _{0i} \left[1+\exp
\left(\frac{r-R_{0i}}{a_{i}}\right)\right]^{-1}, \; \; \;
i=\left\{ {n,p}\right\}.
\end{equation}
For the three quantities $\rho _{0i}$ , $R_{0i}$ and $a_{i}$ in
the equation, only two of them are independent because of the
conservation of particle number $N_i = \int \rho_i({\bf r}) d{\bf
r} $,   $N_i = \left\{ {N,Z}\right\} $. For example, $\rho _{0p}$
can be expressed as a function of $R_{0p}$ and $a_{p}$,
\begin{eqnarray}
\rho _{0p} \simeq Z \left\{ \frac{4}{3}\pi R_{0p}^{3}\left[ 1+\pi
^{2}\left( \frac{a_{p}}{R_{0p}}\right)^{2}\right] \right\}^{-1}
\end{eqnarray}
with high accuracy \cite{Gram82} when $R_{0p} \gg a_{p}$. Here,
$R_{0p}$, $a_{p}$ $R_{0n}$, $a_{n}$ are the radius and diffuseness
for proton and neutron density distributions, respectively.

By using optimization algorithm, one can obtain the minimal energy
$E_{ b}$ as well as the corresponding $R_{0p}$, $a_{p}$, $R_{0n}$,
$a_{n}$ for the neutron and proton density distributions.

The Skyrme force SkM*\cite{Bart82} is adopted in the calculations
since SkM* is very successful for describing the bulk properties
and surface properties of nuclei. It is a well known fact that ETF
calculations with a reasonable effective interaction reproduce
experimental binding energies with a very good accuracy, when
Strutinski-type shell corrections and pairing corrections are
taken into account. And the obtained charge root-mean-square radii
in this work for shell closed nuclei are very close to the
corresponding experimental data \cite{Br84,Fr95}. In addition we
have made a comparison between the results with RDV method and the
Skyrme-HF method and find that the surface diffuseness obtained by
RDV method are little smaller than that by Skyrme-HF calculations.
It is due to the neglect of the higher order term corrections in
the extended Thomas-Fermi (ETF) approach\cite{brack}.  We have
made a check that the surface diffuseness calculated with RDV
method increases if all terms up to fourth order in the spatial
derivatives in ETF energy density functional (ETF4) are taken into
account.

\begin{center}
\textbf{C. the Calculations of Fusion Barriers}
\end{center}
The interaction potential $V_b(R)$ between reaction partners can
be written as
\begin{eqnarray}
V_b(R) = E_{tot}(R) - E_1 - E_2,
\end{eqnarray}
where $R$ is the center-to-center distance between two nuclei, the
$E_{tot}(R)$ is the total energy of the interacting nuclear
system, $E_1$ and $E_2$ are the energies of individual nuclei
(projectile and target), respectively. The interaction potential
$V_b(R)$ is also called entrance-channel potential in
ref.\cite{Deni02} or fusion barrier\cite{Bass74,Dob03}, in the
following we take the term of fusion barrier. The $E_{tot}(R)$,
$E_1$, $E_2$ are calculated with the same energy-density
functional as that is used in the calculations of nuclear
densities,
\begin{eqnarray}
E_{tot}(R) =  \int \; {\mathcal H}[ \rho_{1p}({\bf
r})+\rho_{2p}({\bf r}-{\bf R}), \rho_{1n}({\bf r})+\rho_{2n}({\bf
r}-{\bf R})] \; d{\bf r}, \nonumber
\end{eqnarray}
\begin{eqnarray}
E_1 = \int \; {\mathcal H}[ \rho_{1p}({\bf r}), \rho_{1n}({\bf
r})] \;
d{\bf r}, \\
E_2 = \int \; {\mathcal H}[ \rho_{2p}({\bf r}), \rho_{2n}({\bf
r})] \; d{\bf r}.
\end{eqnarray}
Here,
$\rho_{1p}$, $\rho_{2p}$, $\rho_{1n}$ and $\rho_{2n}$ are the
frozen proton and neutron densities of the projectile and target,
determined in the previous section.

For calculating the nuclear interaction energies and Coulomb
energies, multi-dimensional integral\cite{Bern91} is performed.
For a certain reaction system, the fusion barrier is calculated in
a region from $R=7fm$ to $20fm$ with step of $\Delta R=0.25fm$.
Fig.1 shows the fusion barrier for the reaction
$^{28}$Si+$^{92}$Zr. The solid curve denotes the results based on
the density distributions obtained by RDV method with ETF2. The
dotted curve denotes the proximity potential\cite{Myers00}. The
results with the density determined by restricted density
variational calculations are very close to those of proximity
potential at the region where the densities of two nuclei do not
overlap.

\begin{figure}
\includegraphics[angle=-90,width=0.8\textwidth]{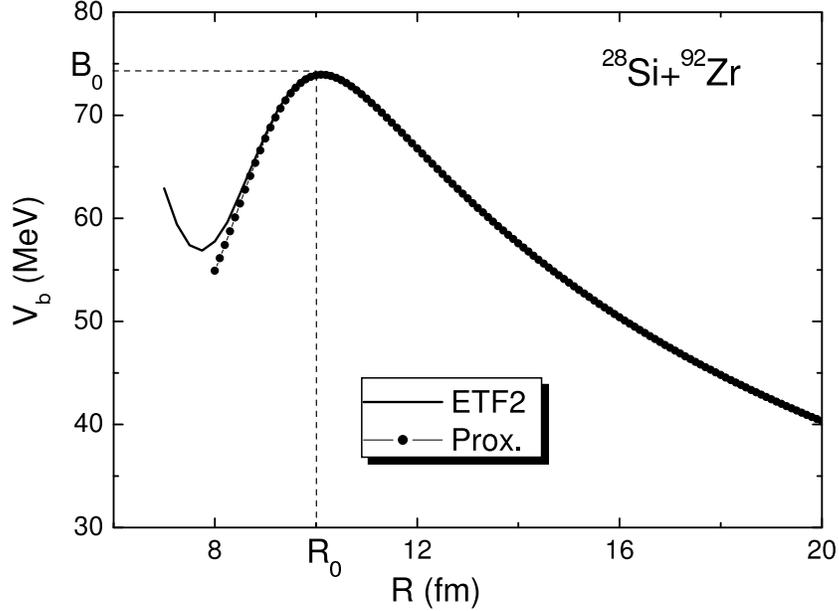}
 \caption{The entrance-channel fusion barrier for
$^{28}$Si+$^{92}$Zr.} \label{fig1}
\end{figure}

The fusion barriers for more than 80 fusion systems ($Z_1Z_2>150$)
are calculated based on ETF2. Part of them are listed in Table.I.
Table.I lists the barrier height $B_0$, radius $R_0$ (see Fig.1)
and the curvature $\hbar\omega_0$ of the barrier for a series of
reactions. Here the curvature of the barrier, $\hbar\omega_0$, is
obtained approximately through fitting the barrier at the region
from $R_0-1.25fm$ to $R_0+1.25fm$ by an inverted parabola. From
the table, one can find that the curvature of the fusion barrier
increases and the Q-value for complete fusion reaction decreases
with the increase of product $Z_1Z_2$.

For exploring the influence of order $\hbar^4$
terms\cite{brack,Bart02}, full fourth order ETF (ETF4) is applied
to calculate both the densities and the fusion barriers for
reactions $^{28}$Si+ $^{28}$Si, $^{40}$Ca+$^{48}$Ca,
$^{16}$O+$^{208}$Pb, and $^{48}$Ca+$^{208}$Pb. The similar work
had been done by A. Dobrowolski et al. in \cite{Dob03}. We find
that when the $\hbar^4$ terms\cite{brack,Bart02} are included in
the determination of the neutron and proton densities, the surface
diffuseness of densities of nuclei increases and thus the height
of fusion barrier calculated is lower (about $1 \sim 2MeV$) than
those with ETF2. The fusion barriers for these reactions
calculated with different approach are shown in Fig.2. The dotted
curves and the solid curves denote the results with ETF2 and ETF4
respectively, and the crossed curves and the dashed curves denote
the results of proximity potential and those of the analytical
form proposed in\cite{Dob03}, respectively. Our calculation
results of barriers with ETF4 are very close to the analytical
form in\cite{Dob03}. We find that the barriers calculated with
ETF2 are more close to the proximity potential for light and
medium-heavy systems (for example $Z_1 Z_2<680$), while the
barrier calculated with ETF4 are more close to those of proximity
potential for heavy systems. It implies that including fourth
order terms of ETF approach is important in improving the surface
diffuseness for heavy nuclei.

\begin{figure}
\includegraphics[angle=-90,width=0.8\textwidth]{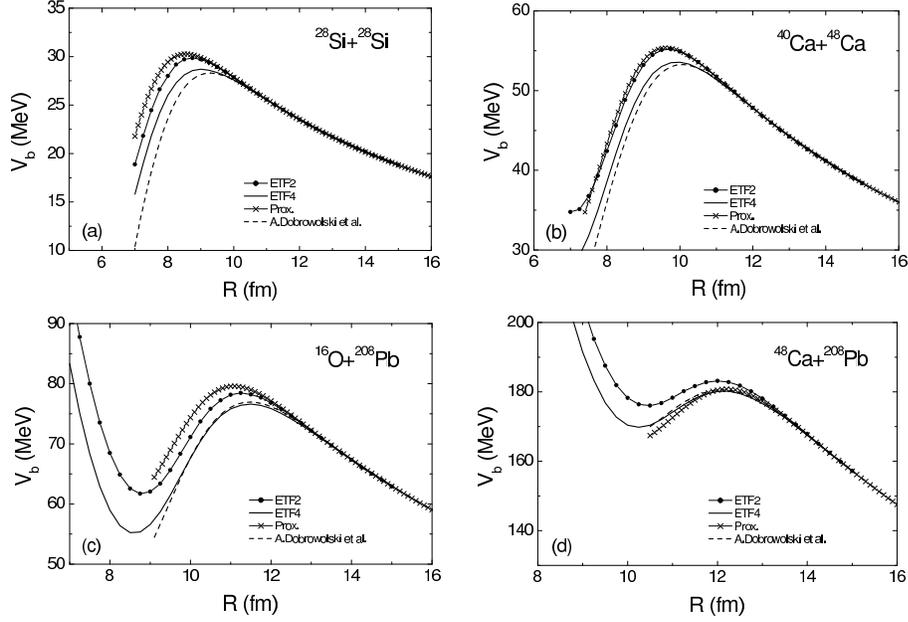}
 \caption{The fusion barriers for $^{28}$Si+$^{28}$Si,
$^{40}$Ca+$^{48}$Ca, $^{16}$O+$^{208}$Pb and $^{48}$Ca+$^{208}$Pb.
The dotted curves and the solid curves denote the results with
ETF2 and ETF4 approach, respectively, and the crossed curves and
the dashed curves denotes the results of proximity potential and
those of the analytical form proposed in\cite{Dob03},
respectively.} \label{fig2}
\end{figure}

Fig.3 shows the barrier heights $B_0$ of the selected systems with
ETF2 approach and the relative deviations of $B_0$ from the
proximity potentials $B_{\rm prox}$\cite{Myers00} as function of
$Z_1Z_2$. The circles and crosses in Fig.3(a) denote the results
of our calculations and those of proximity potential,
respectively. Our calculation results for the barriers are quite
close to the proximity potential and the relative deviations are
less than $2.5\%$ (see Fig.3(b)) generally. From the figure one
can also see that the barrier heights linearly increase with
product $Z_1Z_2$. The solid line in Fig.3(a) is a linear function
fit of the calculation results, i.e.
$B_{fit}=\frac{e^{2}Z_1Z_2}{R_{fit}}+V_0$ with $R_{fit}=13.7 fm$
and $V_0=13.0MeV$.

\begin{figure}
\includegraphics[angle=-90,width=0.8\textwidth]{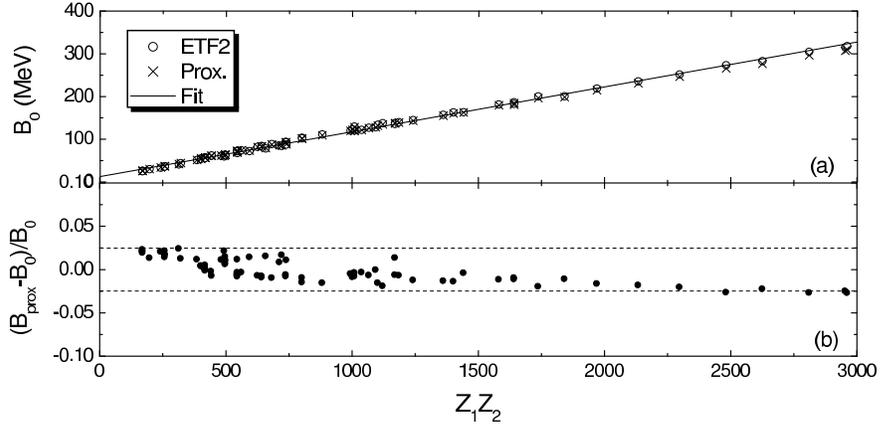}
 \caption{a) The barrier heights $B_{0}$ for a series
of reaction systems and b) The relative deviations from the
results of proximity potential. The circles and crosses in
Fig.3(a) denote the calculation results with RDV and proximity
potential, respectively. The solid line is the linearly fitting to
our calculations.} \label{fig3}
\end{figure}

\begin{table}
\caption{The fusion barriers of a series of reactions with ETF2
approach.}

\begin{tabular}{ccccccc}
\hline\hline
 Reaction & $R_{0}(fm)$  & $ B_{0}(MeV)$  & $ \hbar w_{0}(MeV)$  &
$Z_{1}Z_{2}$  & Q-value$(MeV)$ & Reference \\ \hline
$^{28}$Si+$^{28}$Si & 8.75 & 29.86 & 2.37 & 196 & 10.91 & \cite{1} \\
$^{12}$C+$^{92}$Zr & 9.75 & 33.34 & 2.10 & 240 & 0.94 & \cite{3} \\
$^{16}$O+$^{70}$Ge & 9.25 & 36.52 & 2.84 & 256 & 2.51 & \cite{4} \\
$^{16}$O+$^{72}$Ge & 9.5 & 36.27 & 2.37 & 256 & 6.30 & \cite{4} \\
$^{16}$O+$^{73}$Ge & 9.5 & 36.16 & 2.44 & 256 & 8.84 & \cite{4} \\
$^{16}$O+$^{74}$Ge & 9.5 & 36.08 & 2.50 & 256 & 10.61 & \cite{4} \\
$^{16}$O+$^{76}$Ge & 9.5 & 35.80 & 2.63 & 256 & 10.51 & \cite{4} \\
$^{12}$C+$^{128}$Te & 10.25 & 40.98 & 2.50 & 312 & -0.91 & \cite{5} \\
$^{16}$O+$^{92}$Zr & 9.75 & 43.88 & 2.86 & 320 & -3.94 & \cite{3} \\
$^{16}$O+$^{112}$Cd & 10.0 & 51.12 & 3.19 & 384 & -9.91 & \cite{6} \\
$^{27}$Al+$^{70}$Ge & 9.75 & 57.37 & 3.26 & 416 & -5.17 & \cite{9} \\
$^{27}$Al+$^{72}$Ge & 9.75 & 57.05 & 3.39 & 416 & -4.21 & \cite{9} \\
$^{27}$Al+$^{73}$Ge & 9.75 & 56.87 & 3.46 & 416 & -2.90 & \cite{9} \\
$^{27}$Al+$^{74}$Ge & 9.75 & 56.72 & 3.51 & 416 & -3.21 & \cite{9} \\
$^{27}$Al+$^{76}$Ge & 10.0 & 56.31 & 3.01 & 416 & -2.39 & \cite{9} \\
$^{35}$Cl+$^{54}$Fe & 9.5 & 61.89 & 3.67 & 442 & -17.77 & \cite{7} \\
$^{16}$O+$^{144}$Nd & 10.5 & 61.35 & 3.27 & 480 & -22.43 & \cite{11} \\
$^{12}$C+$^{204}$Pb & 11.0 & 59.80 & 3.19 & 492 & -28.40 & \cite{21} \\
$^{16}$O+$^{144}$Sm & 10.5 & 63.62 & 3.27 & 496 & -28.55 & \cite{13} \\
$^{17}$O+$^{144}$Sm & 10.5 & 63.26 & 3.41 & 496 & -24.43 & \cite{13} \\
$^{16}$O+$^{147}$Sm & 10.5 & 63.31 & 3.37 & 496 & -24.65 & \cite{14} \\
$^{16}$O+$^{148}$Sm & 10.5 & 63.20 & 3.41 & 496 & -23.09 & \cite{13} \\
$^{16}$O+$^{149}$Sm & 10.5 & 63.08 & 3.46 & 496 & -21.71 & \cite{14} \\
$^{16}$O+$^{150}$Sm & 10.5 & 62.96 & 3.49 & 496 & -20.21 & \cite{14} \\
$^{16}$O+$^{154}$Sm & 10.75 & 62.43 & 3.07 & 496 & -16.43 & \cite{13} \\
$^{16}$O+$^{166}$Er & 10.75 & 67.83 & 3.45 & 544 & -25.13 & \cite{16} \\
$^{28}$Si+$^{92}$Zr & 10.0 & 74.20 & 3.97 & 560 & -28.12 & \cite{3} \\

\end{tabular}
\end{table}
\begin{table}
\begin{tabular}{ccccccc}
Reaction & $R_{0}(fm)$ & $B_{0}(MeV)$ & $\hbar w_{0}(MeV)$ &
$Z_{1}Z_{2}$ & Q-value$(MeV)$ & Reference \\ \hline
$^{16}$O+$^{186}$W & 11.0 & 72.26 & 3.52 & 592 & -21.30 & \cite{13} \\
$^{32}$S+$^{89}$Y & 10.25 & 82.31 & 3.76 & 624 & -36.58 & \cite{18} \\
$^{33}$S+$^{90}$Zr & 10.25 & 84.12 & 3.87 & 640 & -39.76 & \cite{19} \\
$^{33}$S+$^{92}$Zr & 10.25 & 83.74 & 3.97 & 640 & -35.51 & \cite{19} \\
$^{16}$O+$^{208}$Pb & 11.25 & 78.46 & 3.54 & 656 & -46.49 & \cite{20} \\
$^{35}$Cl+$^{92}$Zr & 10.25 & 88.61 & 4.14 & 680 & -39.37 & \cite{3}\\
$^{19}$F+$^{197}$Au & 11.25 & 85.21 & 3.69 & 711 & -35.92 & \cite{21} \\
$^{16}$O+$^{232}$Th & 11.5 & 84.47 & 3.68 & 720 & -36.53 & \cite{22,22a} \\
$^{32}$S+$^{110}$Pd & 10.5 & 94.00 & 4.13 & 736 & -35.37 & \cite{23} \\
$^{36}$S+$^{110}$Pd & 10.5 & 92.85 & 4.28 & 736 & -38.01 & \cite{23} \\
$^{19}$F+$^{208}$Pb & 11.25 & 87.44 & 3.87 & 738 & -50.07 & \cite{24} \\
$^{40}$Ca+$^{90}$Zr & 10.25 & 103.74 & 4.31 & 800 & -57.27 & \cite{25} \\
$^{40}$Ca+$^{96}$Zr & 10.5 & 102.22 & 4.28 & 800 & -41.13 & \cite{25} \\
$^{50}$Ti+$^{90}$Zr & 10.5 & 111.16 & 4.49 & 880 & -64.73 & \cite{26} \\
$^{32}$S+$^{154}$Sm & 11.0 & 120.22 & 4.47 & 992 & -60.69 & \cite{27} \\
$^{28}$Si+$^{178}$Hf & 11.25 & 120.65 & 4.41 & 1008 & -64.77 & \cite{29} \\
$^{29}$Si+$^{178}$Hf & 11.25 & 120.31 & 4.42 & 1008 & -65.70 & \cite{29} \\
$^{30}$Si+$^{186}$W & 11.5 & 122.14 & 4.24 & 1036 & -70.22 & \cite{21} \\
$^{31}$P+$^{175}$Lu & 11.25 & 126.97 & 4.46 & 1065 & -70.45 & \cite{29} \\
$^{28}$Si+$^{198}$Pt & 11.5 & 128.04 & 4.38 & 1092 & -78.75 & \cite{31} \\
$^{32}$S+$^{181}$Ta & 11.25 & 138.15 & 4.56 & 1168 & -80.58 & \cite{32} \\
$^{32}$S+$^{182}$W & 11.25 & 140.02 & 4.52 & 1184 & -84.93 & \cite{33} \\
$^{132}$Sn+$^{64}$Ni & 11.5 & 162.75 & 4.61 & 1400 & -111.05 & \cite{34} \\
$^{40}$Ca+$^{208}$Pb & 11.75 & 186.57 & 4.62 & 1640 & -136.69 & \cite{35} \\
$^{48}$Ca+$^{208}$Pb & 12.0 & 183.17 & 4.43 & 1640 & -153.80 & \cite{35} \\
\hline\hline
\end{tabular}
\end{table}

\begin{center}
\textbf{III.  FUSION EXCITATION FUNCTIONS }
\end{center}

\begin{center}
\textbf{A. The Parametrization of Barrier Distribution}
\end{center}
According to Wong's formula\cite{Wong73}, the fusion excitation
function for penetrating a parabolic barrier can be expressed as
\begin{equation}
\sigma _{fus}^{(1)}(E_{\rm c.m.},B_0)=\frac{\hbar \omega
_{0}R_{0}^{2}}{2E_{\rm c.m.}}\ln \left( 1+\exp\left[ \frac{2\pi
}{\hbar \omega _{0}}(E_{\rm c.m.}-B_{0})\right] \right),
\end{equation}
 where $E_{\rm c.m.}$ denotes the center-of-mass energy, $B_0$, $R_0$ and $\hbar\omega_0$
 are the barrier height, radius and curvature, respectively.
This expression is based on the one-dimensional barrier
penetration model. The one-dimensional barrier penetration model
with empirically determined potential parameters is successful in
describing the fusion excitation functions for light systems and
heavy systems at energies above the barrier with some exceptions,
but fails in describing sub-barrier fusion for heavy systems. It
is found that for sub-barrier fusion of heavy systems, the
measured fusion cross sections are of up to several orders of
magnitude higher than the predictions of the model.

\begin{figure}
\includegraphics[angle=-90,width=0.8\textwidth]{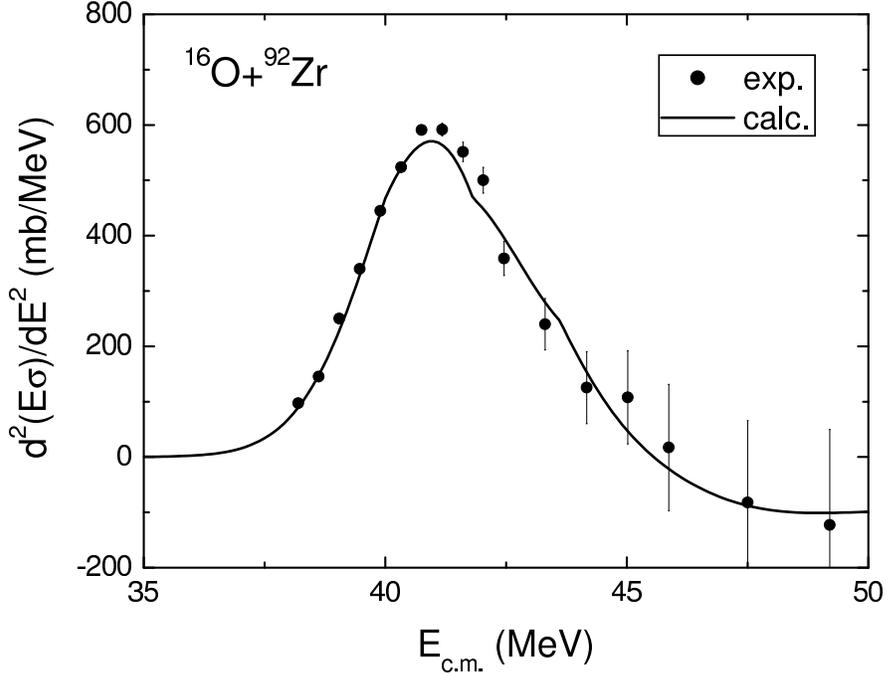}
 \caption{The fusion barrier distribution for
$^{16}$O+$^{92}$Zr. The distribution is evaluated with $\Delta
E_{\rm c.m.}=1.8MeV$. The solid dots and solid curve denote the
experimental data and our calculation results, respectively.}
\label{fig4}
\end{figure}

\begin{figure}
\includegraphics[angle=-90,width=0.8\textwidth]{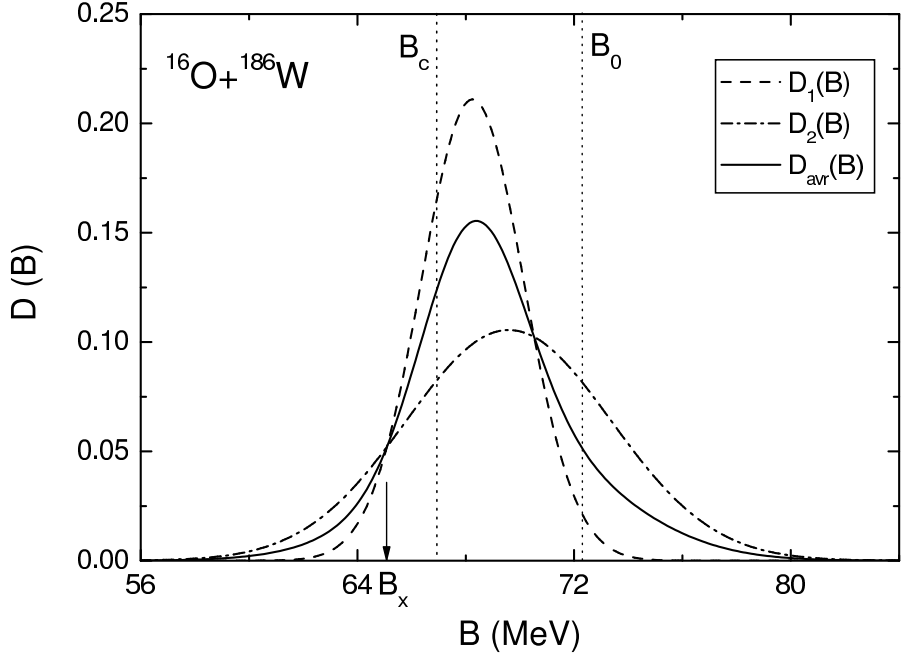}
 \caption{The weighing function for the fusion
reaction of $^{16}$O+$^{186}$W. The dashed and dot-dashed curves
denote Gaussian distributions $D_1(B)$ and $D_2(B)$, respectively.
The solid curve denotes the average value, i.e. $D_{\rm
avr}(B)=[D_1(B)+D_2(B)]/2$. The vertical dotted lines denote the
positions of $B_0$ and $B_c$, respectively. The arrow $B_{x}$
denotes the cross point between the curves $D_1(B)$ and $D_2(B)$
at the left side of the peak of $D_{\rm avr}(B)$.} \label{fig5}
\end{figure}

Taking into account the multi-dimensional character \cite{Stel88}
of the realistic barrier we may introduce the barrier distribution
in order to calculate the total fusion cross sections. Then, the
fusion excitation function in terms of barrier penetration concept
is given by
\begin{eqnarray}
\sigma _{fus}(E_{\rm c.m.})=\int_{0}^{ \infty }D(B) \sigma
_{fus}^{(1)}(E_{\rm c.m.},B)dB,
\end{eqnarray}
 where $\sigma_{fus}^{(1)}(E_{\rm c.m.},B)$ is the fusion excitation
function for the barrier $B$ based on one-dimensional barrier
penetration,  $D(B)$ is a weighing function to describe the
barrier distribution. Here we replace the barrier height $B_0$ in
Wong's formula Eq.(13) by $B$. The distribution function $D(B)$
satisfies
\begin{eqnarray}
\int_{0}^{ \infty }D(B)dB=1.
\end{eqnarray}
$D(B)$ is often taken to be continuous and symmetric distributions
of rectangular or Gaussian shapes\cite{Kra83,Mohan94,Siw04}. The
Wong's formula is a special case when a delta distribution
($D(B)=\delta (B-B_0)$) is taken.

In this work, we propose a parametrization of the weighing
function based on the fusion barrier obtained in the previous
section. That is that we try to propose a macroscopic empirical
barrier distribution rather than explicitly taking into account
the coupling of the fusion motion to internal degrees of freedom
as that is done in the fusion coupled channel model\cite{Hag99}.
Concerning the weighing functions, we first investigate the shape
of barrier distributions extracted from experiments. We find in
most cases the barrier distributions are not symmetric. For
example, in Fig.4 solid circles show the experimental fusion
barrier distribution of $^{16}$O+$^{92}$Zr \cite{Timm98}. One can
see that the fusion barrier distribution of the reaction system is
asymmetric, its left side is steeper than the right side. It
indicates that only one Gaussian distribution is not good enough
to describe the fusion barrier distribution. Motivated by the
shape of the barrier distribution extracted from experiments, we
consider the weighing function to be a superposition of two
Gaussian functions.  Two Gaussian distributions $D_1(B)$ and
$D_2(B)$ are proposed as
\begin{eqnarray}
D_{1}(B)=\frac{\sqrt{\gamma}}{2\sqrt{\pi}w_{1}} \exp \left[ -\gamma \frac{%
(B-B_{1})^{2}}{(2w_{1})^{2}}\right]
\end{eqnarray}
and
\begin{eqnarray}
D_{2}(B)=\frac{1}{2\sqrt{\pi}w_{2}}\exp \left[ -\frac{%
(B-B_{2})^{2}}{(2w_{2})^{2}}\right],
\end{eqnarray}
with
\begin{eqnarray}
w_{1}=\frac{1}{4}(B_0-B_{c}),
\end{eqnarray}
\begin{eqnarray}
w_{2}=\frac{1}{2}(B_0-B_{c}),
\end{eqnarray}
\begin{eqnarray}
B_{1}=B_{c}+w_{1},
\end{eqnarray}
\begin{eqnarray}
B_{2}=B_{c}+w_{2}.
\end{eqnarray}
The barrier height $B_0$, radius $R_0$ and curvature
$\hbar\omega_0$ are obtained from Table.I. The $B_{c}=f B_0$ is
the effective barrier height taking into account the coupling
effects to other degrees of freedom, such as dynamical
deformation, nucleon transfer etc. We take the reducing factor
$f=0.926$. The $\gamma$ in $D_1(B)$ is a factor which influence
the width of the distribution $D_1(B)$. The larger the $\gamma$
value is, the narrower the distribution is. For the fusion
reactions between nuclei with non-closed-shell but near the
$\beta$-stability line, we take $\gamma=1$; for the reactions with
neutron closed-shell nuclei or neutron-rich nuclei the value of
$\gamma$ is calculated by a parameterized formula which will be
discussed in the following sub-section.  From the expression of
(16) and (17) one can find that the peaks and the widths of
$D_1(B)$ and $D_2(B)$ only depend on the height of the fusion
barrier $B_0$ except the $\gamma$ in $D_{1}(B)$. The peaks of two
Gaussian distributions locate at different energies between $B_c$
and $B_0$. For light systems, the width of the distribution
becomes very small and thus it is very close to delta
distribution, in consistent with the one-dimensional penetration
model. Both $D_1(B)$ and $D_2(B)$ satisfy Eq.(15). Further we
introduce $D_{\rm avr}(B)=(D_1(B)+D_2(B))/2$. $D_{\rm avr}(B)$
also satisfies Eq.(15). In Fig.5 we show the parameterized
weighing function of $^{16}$O+$^{186}$W. The dashed and dot-dashed
curves denote Gaussian distributions $D_1(B)$ and $D_2(B)$,
respectively. The solid curve denotes the $D_{\rm avr}(B)$. There
are two crossing points between $D_1(B)$ and $D_2(B)$ (or $D_{\rm
avr}(B)$) at both sides of $D_{\rm avr}(B)$. The left one is
located at $B_{x}$ (see Fig.5). We notice that the left side of
$D_{\rm avr}(B)$ is too flat and does not fit the shape of the
barrier distribution extracted from experimental data well (for
example see the shape shown in Fig.4). So we propose an effective
weighing function
\begin{eqnarray}
D_{\rm eff}(B)= \left\{
\begin{array} {r@{\quad:\quad}l}
D_1(B)& B<B_x \\
D_{\rm avr}(B)& B \ge B_x
\end{array} \right.
\end{eqnarray}
(with  $\int D_{\rm eff}(B)\; dB \approx 1$). We notice that the
distribution $D_1(B)$ plays a main role for sub-barrier fusion and
$D_{\rm avr}(B)$ contributes more to the fusion at energies near
and above the barrier.

With the $D_{\rm eff}(B)$ we can, in principle, calculate the
fusion cross sections by the expression
\begin{eqnarray}
\sigma _{fus}(E_{\rm c.m.})=\int_{0}^{ \infty }D_{\rm eff}(B)\sigma
_{fus}^{(1)}(E_{\rm c.m.},B)dB.
\end{eqnarray}
However, the $D_{\rm eff}(B)$ does not satisfy Eq.(15) exactly
because there is a small difference between $D_1(B)$ and $D_{\rm
avr}(B)$ in the left side of the $B_x$ (see Fig.5). To remedy this
defect, we replace the expression (23) by the the following
expression in the calculation of fusion cross sections
\begin{eqnarray}
\sigma_{fus}(E_{\rm c.m.})=\min[\sigma_1(E_{\rm c.m.}),
\sigma_{\rm avr}(E_{\rm c.m.})].
\end{eqnarray}
Here the $\sigma_1(E_{\rm c.m.})$ and $\sigma_{\rm avr}(E_{\rm
c.m.})$ are calculated by expression (14) with $D(B)=D_{1}(B)$ and
$D(B)=D_{\rm avr}(B)$, respectively. Now the normalization
condition Eq.(15) is always satisfied for the weighing function
adopted in the calculations of fusion cross sections at each
$E_{\rm c.m.}$. We have checked that the fusion cross sections
calculated with the expression (24) are very close to those
calculated with the expression (23) when the $\gamma$ is not too
large. In fact, we find that the results calculated with Eq.(24)
is in better agreement with experimental data. To check the
barrier distribution obtained with our model, we also show the
barrier distribution obtained from the calculation in Fig.4 by the
solid curve. The good agreement between experimental data and
calculation results indicates the parametrization of the weighing
function is quite reasonable.

\begin{figure}
\includegraphics[angle=-90,width=0.8\textwidth]{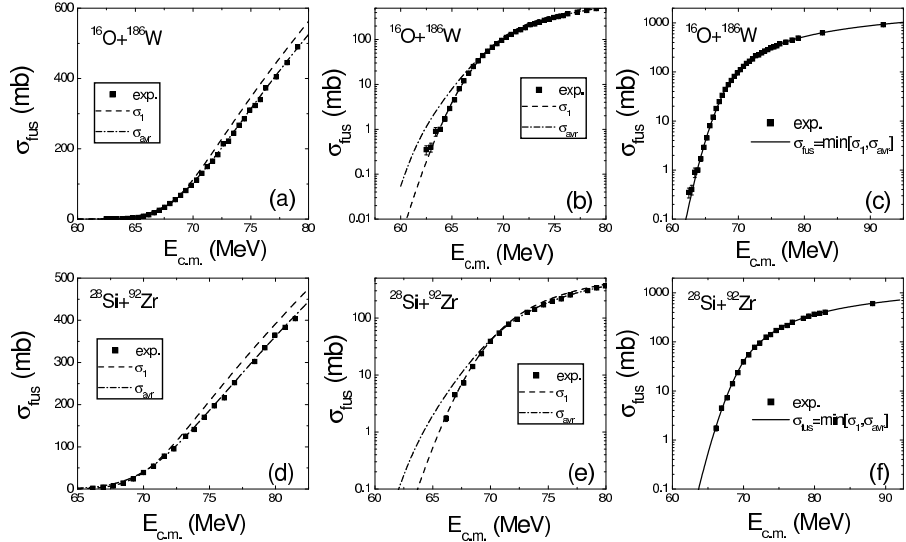}
 \caption{The fusion excitation functions for
$^{16}$O+$^{186}$W and $^{28}$Si+$^{92}$Zr. The dashed and the
dot-dashed curves denote the cross sections $\sigma_1(E_{\rm
c.m.})$ and $\sigma_{\rm avr}(E_{\rm c.m.})$ calculated with
$D(B)=D_1(B)$ and $D(B)=D_{\rm avr}(B)$, respectively.}
\label{fig6}
\end{figure}

\begin{figure}
\includegraphics[angle=-90,width=0.8\textwidth]{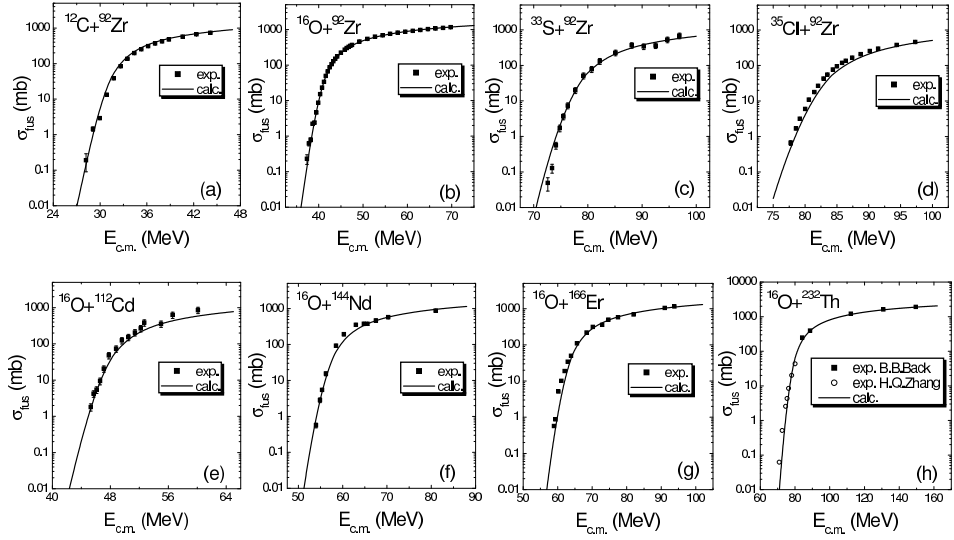}
 \caption{The fusion excitation functions for
$^{12}$C, $^{16}$O, $^{33}$S, $^{35}$Cl+$^{92}$Zr and
$^{16}$O+$^{112}$Cd, $^{144}$Nd, $^{166}$Er, $^{232}$Th. The
squares denote the experimental data and the solid curves denote
the calculation results with $\gamma=1$.} \label{fig7}
\end{figure}

\begin{figure}
\includegraphics[angle=-90,width=0.8\textwidth]{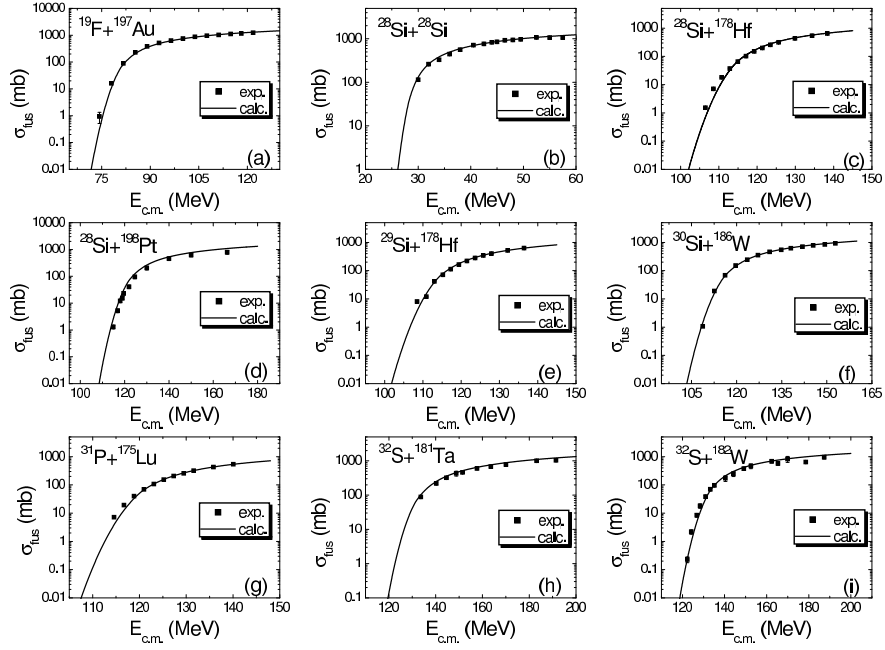}
 \caption{The fusion excitation functions for
$^{19}$F+$^{197}$Au,$^{28}$Si+$^{28}$Si, $^{28,29}$Si+$^{178}$Hf,
$^{28}$Si+$^{198}$Pt, $^{28}$Si+$^{186}$W, $^{31}$P+$^{175}$Lu,
$^{32}$S+$^{181}$Ta and $^{32}$S+$^{182}$W. The squares denote the
experimental data and the solid curves denote the calculation
results with $\gamma=1$.} \label{fig8}
\end{figure}

\begin{figure}
\includegraphics[angle=-90,width=0.8\textwidth]{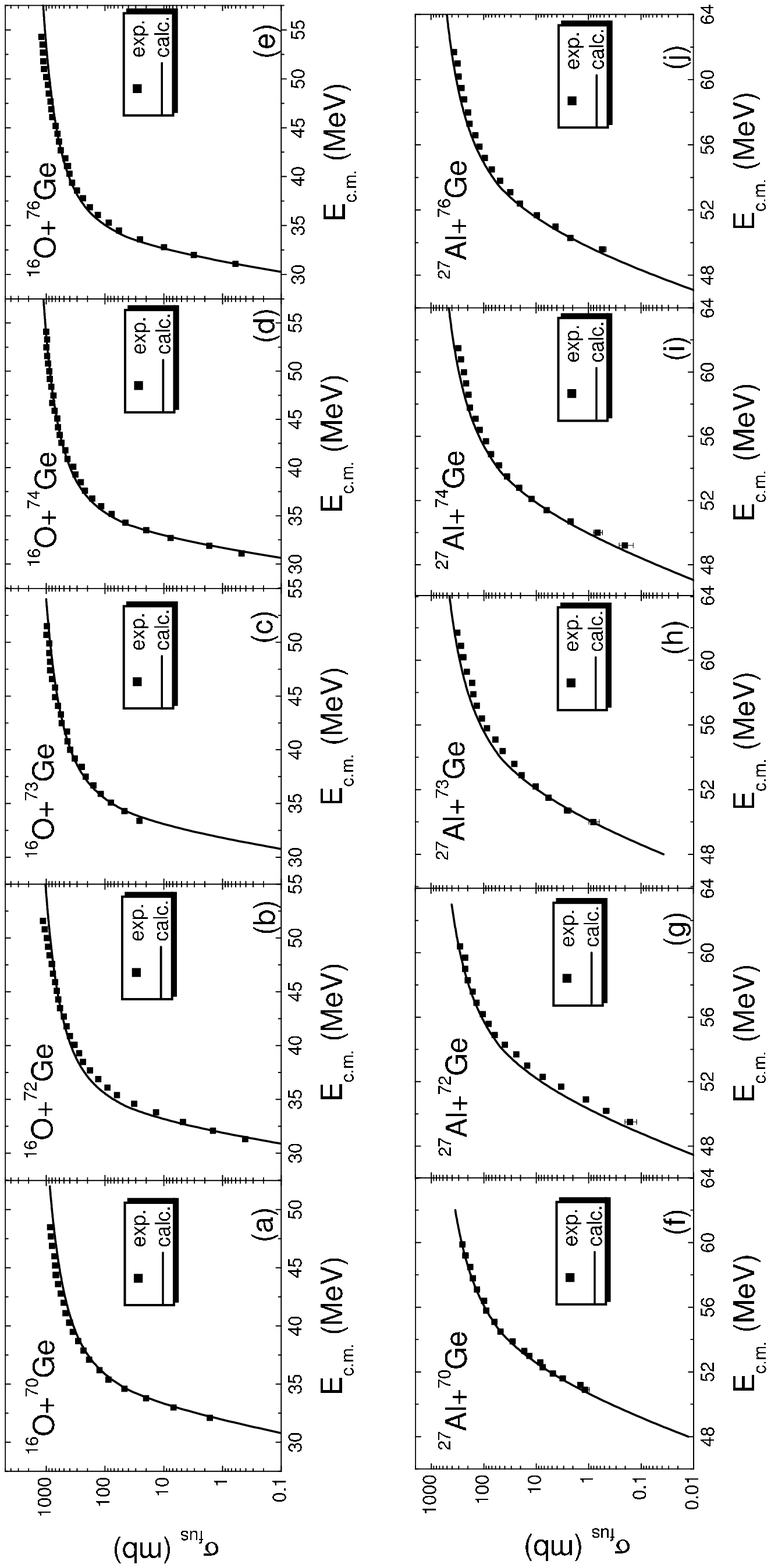}
 \caption{The fusion excitation functions for
$^{16}$O,$^{27}$Al+$^{70,72,73,74,76}$Ge. The squares denote the
experimental data and the solid curves denote the calculation
results with $\gamma=1$. } \label{fig9}
\end{figure}

\begin{center}
\textbf{B. Fusion Reactions between Nuclei with Non-closed-shell
but near  $\beta$-Stability Line}
\end{center}
In this sub-section, the fusion excitation functions for about 30
reaction systems are investigated. For these reaction systems, the
projectile and target nuclei are non-closed-shell but near the
$\beta$-stability line (The shell-closure effects of $^{16}$O are
neglected in this work). The shell effect and excess neutron
effect are weak for this kind reaction systems. Fig.6 shows the
fusion excitation functions of $^{16}$O+$^{186}$W and
$^{28}$Si+$^{92}$Zr. From the figure one can find that at energies
above the barrier $\sigma_{\rm avr}$ (dot-dashed curves) is more
close to the experimental data, while at energies below the
barrier, $\sigma_1$ (dashed curves) is preferable. From Fig.6(c)
and Fig.6(f) one can see that all experimental data spanning the
fusion barrier for both $^{16}$O+$^{186}$W and $^{28}$Si+$^{92}$Zr
can be reproduced well. Fig.7 to Fig.9 show more calculation
results and experimental data for comparison for this kind fusion
reactions. All the fusion excitation functions can be reproduced
very well, which indicates our parametrization of $D_1(B)$ and
$D_2(B)$ is quite useful and reasonable.

\begin{table}
\caption{The $\Delta Q$ and $\gamma$ values for a series reactions
with neutron closed-shell nuclei or neutron-rich nuclei. }
\begin{tabular}{cccc}
\hline\hline
Reaction & $\Delta$Q$(MeV)$ $\;\;\;$ & $\gamma $ $\;\;\;$ & Reference\\
\hline
$^{33}$S+$^{90}$Zr & -4.25 & 3.6 & \cite{19}\\
$^{50}$Ti+$^{90}$Zr & -3.25 & 3.6 &\cite{26}\\
$^{35}$Cl+$^{54}$Fe & -6.58 & 4.8 &\cite{7}\\
$^{16}$O+$^{208}$Pb & 0 & 1.5 &\cite{20}\\
$^{19}$F+$^{208}$Pb & 0 & 1.5 &\cite{21}\\
$^{17}$O+$^{144}$Sm & -0.21 & 1.6& \cite{13}\\
\hline
$^{32}$S+$^{154}$Sm & 9.80 & 0.5 &\cite{27}\\
$^{32}$S+$^{110}$Pd & 4.13 & 0.6 &\cite{23}\\
$^{40}$Ca+$^{96}$Zr & 10.00 & 0.5 &\cite{25}\\
$^{132}$Sn+$^{64}$Ni & 7.50 & 0.7 & \cite{34}\\
\hline
$^{16}$O+$^{144}$Sm & -3.91 & 3.5 &\cite{13}\\
$^{16}$O+$^{147}$Sm & 0 & 1.0 &\cite{14}\\
$^{16}$O+$^{148}$Sm & 1.56 & 0.8& \cite{13}\\
$^{16}$O+$^{149}$Sm & 2.94 & 0.7 &\cite{14}\\
$^{16}$O+$^{150}$Sm & 4.44 & 0.6 &\cite{14}\\
$^{16}$O+$^{154}$Sm & 8.22 & 0.5 &\cite{13}\\
\hline
$^{16}$O+$^{144}$Sm & -3.91 & 3.5 &\cite{13}\\
$^{16}$O+$^{154}$Sm & 8.22 & 0.5 &\cite{13}\\
$^{28}$Si+$^{58}$Ni & -6.81 & 4.4 &\cite{36}\\
$^{28}$Si+$^{64}$Ni & 8.56 & 0.5 &\cite{36}\\
$^{36}$S+$^{90}$Zr & -1.25 & 2.6 &\cite{37}\\
$^{36}$S +$^{96}$Zr & 7.85 & 0.7 &\cite{37}\\
$^{40}$Ca+$^{90}$Zr & -6.14 & 5.1 &\cite{25}\\
$^{40}$Ca+$^{96}$Zr & 10.00 & 0.5 &\cite{25}\\
$^{32}$S+$^{110}$Pd & 4.13 & 0.6 &\cite{23}\\
$^{36}$S+$^{110}$Pd & 1.49 & 1.3 &\cite{23}\\
$^{40}$Ca+$^{48}$Ca & 10.64 & 0.9 & \cite{8}\\
$^{48}$Ca+$^{48}$Ca & 3.09 & 1.7 & \cite{8}\\
 \hline\hline
\end{tabular}
\end{table}

\newpage

\begin{center}
\textbf{C. Fusion Reactions between Nuclei with
Neutron-Shell-Closure or Neutron-rich Nuclei}
\end{center}
The shell effect for neutron closed-shell nuclei and the excess
neutron effect for neutron-rich nuclei play an important role in
the fusion reactions at energies below the fusion barrier.
Fig.10(a) shows the fusion excitation function of
$^{33}$S+$^{90}$Zr. The dashed curve denotes the calculation
results with $\gamma=1$ and the squares denote the experimental
data. One can find that at the energies near and above the barrier
$B_{0}$ the calculation results with $\gamma=1$ are in good
agreement with the experimental data. However, at sub-barrier
energies the fusion cross sections are over-predicted with
$\gamma=1$. Through adjusting the $\gamma$ value, we find that the
fusion excitation function at sub-barrier energies for
$^{33}$S+$^{90}$Zr can be reproduced reasonably well when
$\gamma=3.6$. Fig.10(b) shows the fusion excitation function of
the neutron-rich nuclear fusion reaction $^{32}$S+$^{154}$Sm. For
this reaction, the fusion cross sections at sub-barrier energies
are under-predicted when $\gamma=1$. For this case, a small
$\gamma$ should be taken so that the distribution $D_1(B)$ is
broaden and thus the fusion cross sections are enhanced at
sub-barrier energies. We find that when $\gamma=0.5$ the
sub-barrier fusion cross sections can be reproduced well (see
Fig.10(b)). Through systematically analyzing the fusion excitation
functions of reactions between nuclei with neutron closed-shell
but near the $\beta$-stability line and those with neutron-rich
nuclei, we find that fusion cross sections are suppressed for the
former cases and enhanced for the latter cases compared to the
calculation results with $\gamma=1$. The enhancement of fusion
cross sections for reactions with neutron-rich nuclei compared
with non-neutron-rich nuclei has been found in refs.
\cite{Stel88,Timm98,Wang02}. It is attributed to the neutron
transfer and neck formation which lower the fusion barrier and
thus enhance the fusion cross sections at sub-barrier
energies\cite{Stel88,Wang02}. For neutron closed-shell nuclei, the
strong shell effect suppresses the lowering barrier effect. Based
on above discussion we propose an empirical formula for the
$\gamma$ values used in the weighing function $D_1(B)$ for systems
with the same $Z_{1}$ and $Z_{2}$,
\begin{eqnarray}
\gamma=1-c_0 \Delta Q + 0.5(\delta_{n}^{prog}+\delta_{n}^{targ}),
\end{eqnarray}
where $\Delta Q=Q-Q_0$ denotes the difference between the Q-value
of the system under considering for complete fusion and that of
the reference system. The $Q_0$ is the Q-value of reference
system. $c_0=0.5 MeV^{-1}$ for $\Delta Q<0$ cases and
$c_0=0.1MeV^{-1}$ for $\Delta Q>0$ cases.
$\delta_{n}^{proj(targ)}=1$ for neutron closed-shell projectile
(target) nucleus and $\delta_{n}^{proj(targ)}=0$ for non-closed
cases (The shell-closure effects of $^{16}$O are neglected in this
work as mentioned in the above sub-section). In addition, we
introduce a truncation for $\gamma$ value, i.e. $\gamma$ should
not be smaller than 0.5.

\begin{figure}
\includegraphics[angle=-90,width=0.8\textwidth]{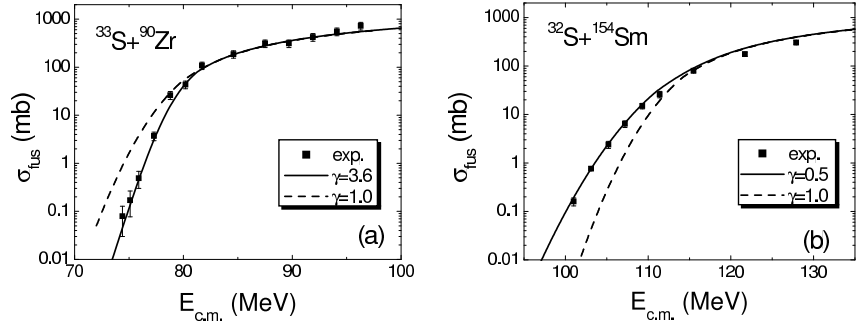}
 \caption{ The fusion excitation functions for
reactions $^{33}$S+$^{90}$Zr, $^{32}$S+$^{154}$Sm. The squares
denote the experimental data. The solid and dashed curves denote
the calculation results with the $\gamma$ values obtained by
formula (25) and with $\gamma=1$, respectively.} \label{fig10}
\end{figure}

The reference system is chosen to be the reaction system with
nuclei along the $\beta$-stability line. More precisely, we do it
as follows: from the periodic table we find the relative atomic
masses of corresponding elements of projectile and target ($M_{\rm
a.m.}^{proj}$, $M_{\rm a.m.}^{targ}$), then the mass numbers for
reaction partners of reference system ($A_{0}^{proj}$ and
$A_{0}^{targ}$) can be obtained by inequality $A_{0}^{i}-1 <
M_{\rm a.m.}^{i} \le A_{0}^{i} $. $i$ denotes projectile
($i=proj$) or target ($i=targ$) nuclei. For example, for three
fusion reactions $^{16}$O+$^{90,92,96}$Zr, $^{16}$O+$^{92}$Zr
reaction system is taken to be the reference system according to
the atomic mass of O and Zr obtained from periodic table and
inequality given above. If there exist certain reaction system for
which the experimental data of fusion excitation function can be
reproduced well by the calculations with $\gamma=1$, then this
reaction system is preferably chosen as the reference system for
the series of reactions with the same $Z_{1}$ and $Z_{2}$. For
example, for the series of reactions of $^{16}$O+Sm, the reaction
$^{16}$O+$^{147}$Sm is taken as the reference system because the
experimental data can be reproduced by the calculations with
$\gamma=1$ (see the following discussion). For this kind of
reactions, there may exist possibility of more than one such
reaction systems for which the experimental data can be described
by the calculations with $\gamma=1$. This situation might be rare
and from our investigation we have not encountered. If this
situation occurs we prefer to choosing the reference system
according to the inequality given above. The $\gamma$ values for
some fusion reactions with neutron closed-shell nuclei or
neutron-rich nuclei are calculated by expression (25) and the
results are listed in Table.II. For fusion reactions between
nuclei with neutron closed-shell and near the $\beta$-stability
line, the $\gamma$ values are larger than 1 (see Table.II), which
means that the width of the barrier distribution $D_1(B)$ becomes
narrow and thus the fusion cross sections of these systems at
sub-barrier energies are suppressed compared with the reference
system. Fig.11 shows the fusion excitation functions of these
fusion reactions. One can see from the figure that the
experimental data are nicely reproduced.

\begin{figure}
\includegraphics[angle=-90,width=0.8\textwidth]{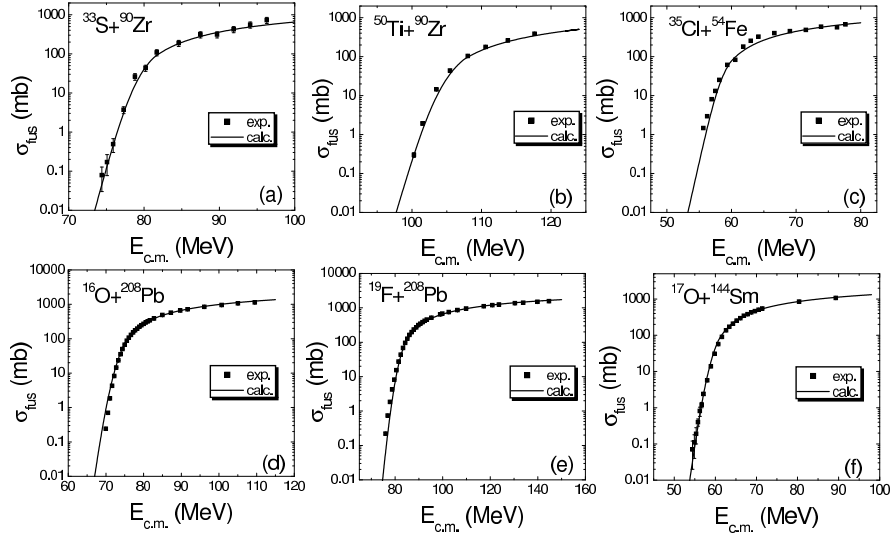}
 \caption{The fusion excitation functions for
reactions $^{33}$S,$^{50}$Ti+$^{90}$Zr, $^{35}$Cl+$^{54}$Fe,
$^{16}$O,$^{19}$F+$^{208}$Pb and $^{17}$O+$^{144}$Sm with
neutron-shell-closure nuclei near the $\beta$-stability line. The
squares denote the experimental data. The solid denote the
calculation results with the $\gamma$ values obtained by formula
(25).} \label{fig11}
\end{figure}

\begin{figure}
\includegraphics[angle=-90,width=0.8\textwidth]{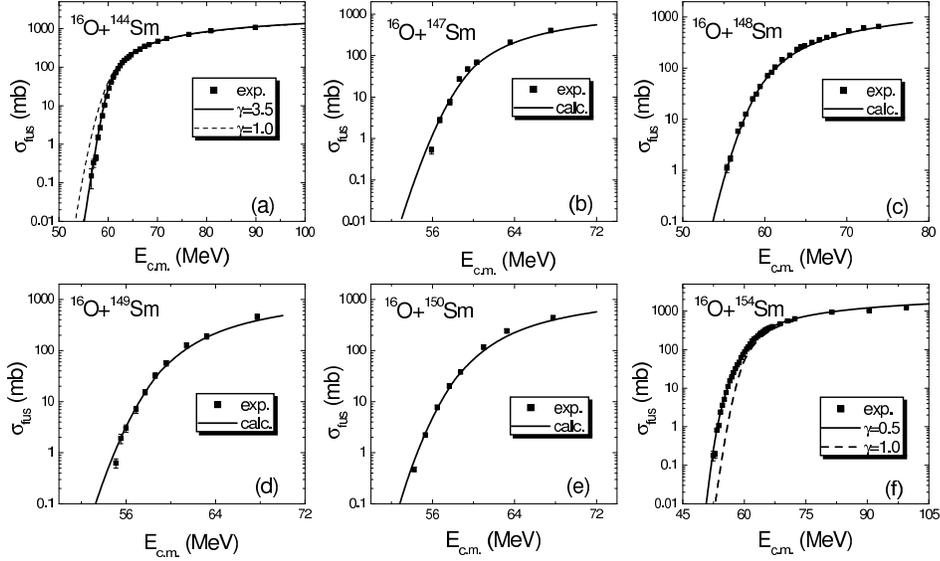}
 \caption{The fusion excitation functions for
$^{16}$O+$^{144,147,148,149,150,154}$Sm. The squares denote the
experimental data. The solid and dashed curves denote the
calculation results with $\gamma$ obtained by formula (25) and
with $\gamma=1$, respectively. } \label{fig12}
\end{figure}

\begin{figure}
\includegraphics[angle=-90,width=0.8\textwidth]{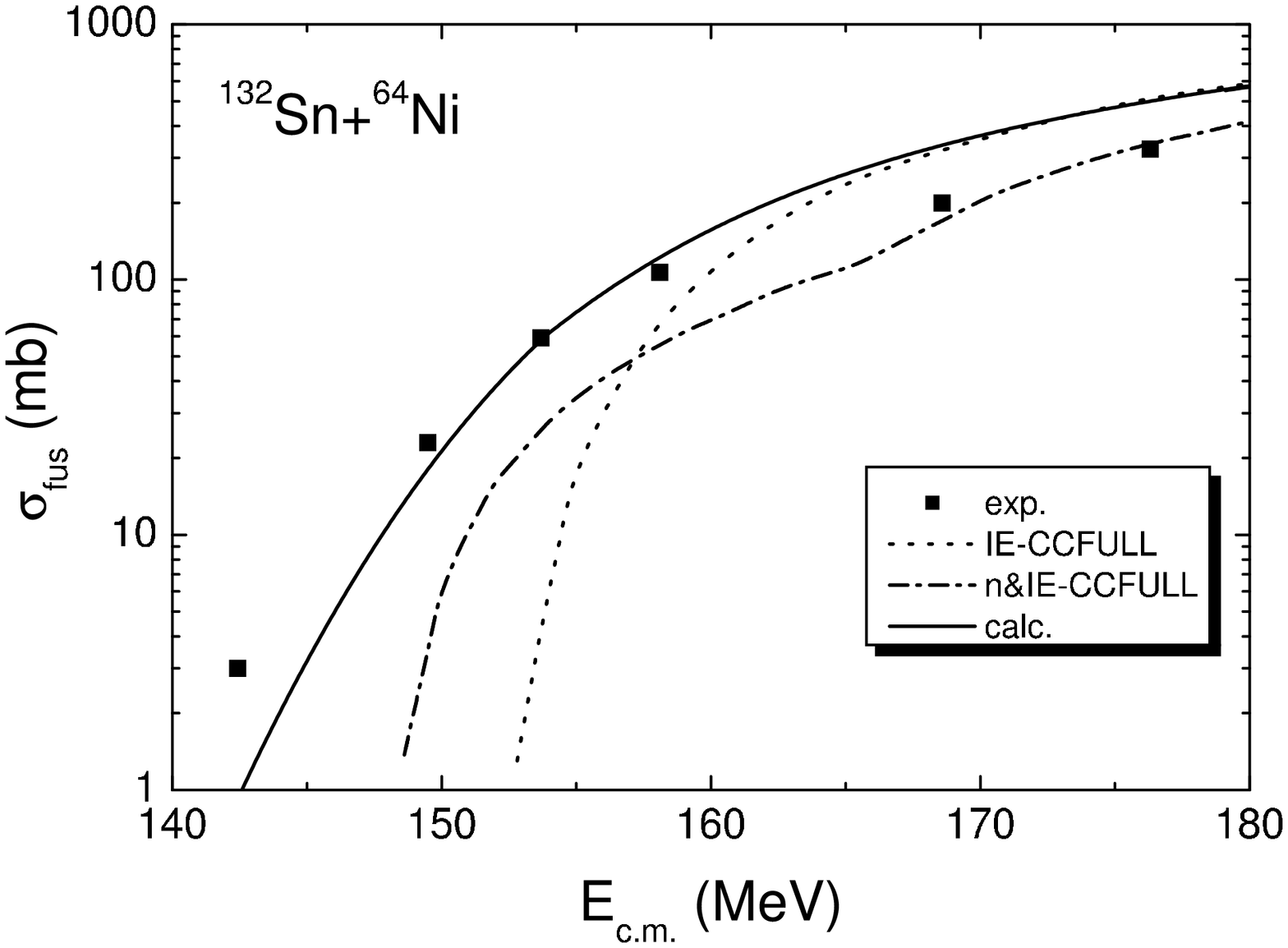}
 \caption{The fusion excitation functions for
$^{132}$Sn+$^{64}$Ni. The squares denote the experimental data and
the solid  curves  denote the calculated results with  $\gamma$
obtained by formula (25). The dot-dashed and doted curves denote
the results of fusion coupled channel model\cite{Liang03},
respectively.} \label{fig13}
\end{figure}

Further, we find that at energies near and above the fusion
barrier, calculated fusion cross sections are not sensitive to the
value of $\gamma$ (see Fig.10 and Fig.12(a) and Fig.12(f)), which
implies that we can calculate the fusion cross sections at
energies near and above the fusion barrier for unmeasured fusion
system by simply taking $\gamma=1$ in our parametrization of
weighing function. For sub-barrier fusion, more sophisticated
investigation for $\gamma$ value is required. For further
investigating the influence of $\gamma$ value, a series of fusion
reactions with $^{16}$O projectile on Sm isotope targets from
$^{144}$Sm to $^{154}$Sm are studied. Fig.12 shows the fusion
excitation functions for $^{16}$O +$^{144-154}$Sm. In the
calculations we find that the fusion cross sections for
$^{16}$O+$^{147}$Sm can be reproduced well by the calculations
with $\gamma=1$ (see Fig.12(b)), so this system is taken as a
reference system for studying other $^{16}$O+$^{144-154}$Sm. The
$\Delta Q$ for $^{16}$O+$^{144}$Sm is negative as obtained from
Table.II and consequently the corresponding $\gamma$ value
calculated by expression (25) is larger than 1. Thus, this fusion
process is unfavored and the fusion cross sections at sub-barrier
energies are suppressed compared with reference system
$^{16}$O+$^{147}$Sm. With the increase in numbers of neutrons, the
Q-values of reaction under consideration increase gradually and
the corresponding $\gamma$ values decrease, which indicates that
the sub-barrier fusion cross sections change from suppression to
enhancement compared with the reference system. Our calculated
results for O+Sm reactions are in good agreement with the
experimental data. For investigating the competition between
suppression and enhancement effect of fusion cross sections, six
pair reactions of $^{16}$O+$^{144,154}$Sm, $^{28}$Si+$^{58,64}$Ni,
$^{36}$S+$^{90,96}$Zr, $^{40}$Ca+$^{90,96}$Zr,
$^{32,36}$S+$^{110}$Pd and $^{40,48}$Ca+$^{48}$Ca are studied. We
find that those reactions with $\gamma <1$ are all enhanced at
sub-barrier energies compared with non-neutron-rich systems, while
those reactions with $\gamma>1$ are all suppressed at sub-barrier
fusion compared with neutron non-closed-shell systems.

For radioactive beam fusion reactions $^{132}$Sn+$^{64}$Ni, the
competition exists since $^{132}$Sn is both neutron-rich and
neutron-shell-closure ($N=82$, $Z=50$). The $\gamma$ value for
this reaction is smaller than 1 (see Table.II) and the enhancement
of the fusion cross sections at sub-barrier energies is expected.
The fusion cross sections of this system are shown in Fig.13.
Squares and solid curve denote the experimental data\cite{34} and
our calculation results, respectively. The results of fusion
coupled-channel model (The dotted curve denotes inelastic
excitations  and the dashed curve denotes both the inelastic
excitations and neutron transfer are considered\cite{34}) are also
presented for comparison. From the comparison one can find that
the agreement in sub-barrier fusion cross sections calculated with
our approach is better than the fusion coupled channel
model\cite{Hag99} calculations. This seems to indicate that there
are still some physical aspects missing in present coupled channel
calculations.

\begin{center}
\textbf{IV. CONCLUSION AND DISCUSSION}
\end{center}

In this work, the Skyrme energy density functional has been
applied to study heavy-ion fusion reactions. The properties of
ground state nuclei are studied by using the restricted density
variational method with the Skyrme energy density functional plus
the semi-classical extended Thomas-Fermi approach (up to second
order in $\hbar$). With the proton and neutron density
distributions obtained in this way, the fusion barriers of a
series of reaction systems are calculated by the same Skyrme
energy density functional. We propose a parametrization for the
weighing function describing the empirical barrier distribution
based on the fusion barrier calculated with Skyrme energy density
functional.  The weighing functions of the barrier are assumed to
be the superposition of two Gaussian functions. With the
parametrization of the weighing function for the empirical barrier
distribution, fusion excitation functions for more than 50 systems
are calculated. A large number of measured fusion excitation
functions spanning the fusion barriers can be reproduced well. The
competition between suppression and enhancement effects on
sub-barrier fusion caused by neutron-shell-closure and excess
neutron effects have been investigated.

However, the proton-shell-closure effects, dynamical effects
\cite{Wang02,Wang04} as well as the effects due to large
deformation of nuclei in the fusion reactions have not been taken
into account in present work, yet. All those effects are very
important in fusion dynamics but they are beyond the scope of this
work.  The study on these aspects is under way.

\section{Acknowledgements}

This work is supported by China Postdoctoral Science Foundation,
and National Natural Science Foundation of China, No. 10235030,
10235020, 10347142, 10375001, Major State Basic Research
Development Program under Contract No. G20000774, CAS-grant
KJCX2-SW-N02. The website
(http://nrv.jinr.ru/nrv/webnrv/fusion/reactions.php), where
experimental data for some fusion reactions are accumulated, is
acknowledged.


\newpage

\begin{thebibliography}{99}


\bibitem{3} \textsf{ J.O. Newton, C.R. Morton and M. Dasgupta,
Phys. Rev. C \textbf{64}, 64608 (2001).}

\bibitem{13} \textsf{J.R. Leigh, M. Dasgupta and D.J. Hinde,
Phys. Rev. C \textbf{52}, 3151 (1995).}

\bibitem{20} \textsf{C.R. Morton, A.C. Berriman and M.
Dasgupta, Phys. Rev. C \textbf{60}, 044608 (1999).}

\bibitem{New04} \textsf{J. O. Newton, R. D. Butt, M. Dasgupta et al., Phys. Rev. C
\textbf{70}, 024605 (2004).}

\bibitem{Zag02} \textsf{V. I. Zagrebaev, et.al., Phys. Rev. C\textbf{65}, 014607(2002).
}
\bibitem{Zag03} \textsf{V. I. Zagrebaev, et.al., Phys. Rev. C\textbf{67}, 061601(R)(2003).
}
\bibitem{Stel88}  \textsf{P.H.Stelson, Phys. Lett. B \textbf{205},190
(1988).}

\bibitem{Deni02} \textsf{V.Yu. Denisov and W. Noerenberg, Eur. Phys. J. \textbf{A15}, 375 (2002).}

\bibitem{Vau72} \textsf{D. Vautherin, D.M. Brink, Phys. Rev. C \textbf{5}, 626 (1972). }

\bibitem{brack} \textsf{M. Brack, C. Guet, H.-B. Hakanson,
Phys. Rep. \textbf{123}, 275 (1985).}

\bibitem{Bart02} \textsf{J. Bartel and K. Bencheikh, Eur. Phys. J, \textbf{A14}, 179 (2002).}


\bibitem{Hoh64} \textsf{P. Hohenberg, W. Kohn, Phys. Rev. \textbf{136}, B864(1964).}

\bibitem{Bart85} \textsf{ J. Bartel, M. Brack and M. Durand, Nucl. Phys. A  \textbf{445}, 263 (1985).}

\bibitem{Cen90}  \textsf{ M. Centelles, M. Pi, X. Vinas, F. Garcias, M. Barranco,
                  Nucl. Phys. A \textbf{510}, 397 (1990).}

\bibitem{Bon87}  \textsf{P. Bonche, H. Flocard and P. H. Heenen, Nucl. Phys. A\textbf{467},
 115 (1987).}

\bibitem{Gram82} \textsf{B. Grammaticos, Z. Phys. A\textbf{305} 257 (1982).}

\bibitem{Bart82} \textsf{J. Bartel, Ph. Quentin, M. Brack, C. Guet and H.B.
Hakansson, Nucl. Phys. A \textbf{386}, 79 (1982). }

\bibitem{Audi95} \textsf{ G. Audi and A. H. Wapstra, Nucl. Phys. A\textbf{595}, 409 (1995).}

\bibitem{Br84} \textsf{B.A. Brawn, et.al., J. Phys. G: Nucl. Phys. \textbf{10}, 1683 (1984) and refrences therein.}

\bibitem{Fr95} \textsf{G. Fricke, et. al. Nucl. Data Table \textbf{60}, 177 (1995). }

\bibitem{Bass74}  \textsf{R.Bass, Nucl. Phys. A \textbf{231}, 45 (1974).}

\bibitem{Dob03} \textsf{A. Dobrowolski, K. Pomorski and J. Bartel, Nucl. Phys. A \textbf{729} 713
(2003).}

\bibitem{Bern91}  \textsf{J. Berntsen, T. O. Espelid and
A. Genz, ACM Trans. Math. Softw., \textbf{17}, No.4, 452 (1991).}

\bibitem{Bein75}  \textsf{M. Beiner, H. Flocard, et al., Nucl.
Phys. A \textbf{238} 29 (1975).}

\bibitem{Myers00} \textsf{W. D. Myers, W. J. Swiatecki, Phys. Rev. C\textbf{62}, 044610
(2000). }

\bibitem{Wong73}  \textsf{C.Y.Wong, Rev. Lett. \textbf{31}, 766 (1973).}

\bibitem{Kra83}  \textsf{H.J.Krappe, et al. Z. Phys. A
\textbf{314}, 23 (1983).}

\bibitem{Mohan94}  \textsf{A.K.Mohanty, S.k.Kataria, Pramana, \textbf{43}
No.4,  319 (1994).}

\bibitem{Siw04} \textsf{K.Siwek-Wilczynska and J.Wilczynski, Phys. Rev. C\textbf{69},
024611 (2004). }


\bibitem{Hag99} \textsf{K. Hagino, N. Rowley, and A. T. Kruppa, Comput. Phys. Commun. \textbf{123}, 143
 (1999).}


\bibitem{Wang02} \textsf{Ning Wang, Zhuxia Li, Xizhen Wu, Phys. Rev. C \textbf{65}, 064608 (2002). }


\bibitem{Timm98} \textsf{H. Timmers, D. Ackermann, et al., Nucl. Phys.
\textbf{A 633}, 421 (1998); and the doctoral thesis of H.
Timmers.}


\bibitem{Beck88} \textsf{M.Beckerman, Rep. Prog. Phys. \textbf{51}, 1047 (1988).}

\bibitem{Trot01} \textsf{M. Trotta, et al., Phys. Rev. C \textbf{65},
011601 (2001).}

\bibitem{Liang03} \textsf{J.F.Liang, D.Shapira, C.J.Geene, et al.,Phys. Rev.
Lett. \textbf{91}, 152701 (2003), and references therein. }


\bibitem{Ada04} \textsf{G.G.Adamian, N.V.Antonenko, W.Scheid, Phys. Rev. C \textbf{69}, 044601
(2004). }


\bibitem{1} \textsf{Y. Nagashima, J. Schimizu and T. Nakagava,
Phys. Rev. C \textbf{33}, 176 (1985).}


\bibitem{4} \textsf{ E.F. Aguilera, J.J. Kolata and R.J. Tighe,
Phys. Rev. C \textbf{52}, 3101 (1995).}

\bibitem{5} \textsf{ D.E DiGregorio, Y. Chan and E. Chavez,
Phys. Rev. C \textbf{43}, 687 (1991).}

\bibitem{6} \textsf{D. Ackermann, L.Corradi and D.R. Napoli,
Nucl. Phys. A \textbf{575}, 374 (1994).}

\bibitem{7} \textsf{ E.M. Szanto, R. Liguori Neto and M.C.S.
Figueira, Phys. Rev. C \textbf{41}, 2164 (1990).}

\bibitem{8} \textsf{M. Trotta, A.M. Stefanini and L. Corradi,
Phys. Rev. C \textbf{65}, 011601 (2001).}

\bibitem{9} \textsf{ E.F. Aguilera, J.J. Vega and J.J. Kolata.
Corradi, Phys. Rev. C \textbf{41}, 910 (1990).}

\bibitem{10} \textsf{ J.D. Bierman, P. Chan, M.P. Kelly, J.F.
Liang, A.A. Sonzogni and R. Vandenbosch, Phys. Lab. Annu. Rep.,
University of Washington (1995) pp.19.}

\bibitem{11} \textsf{ G. Duchene, P. Romain and F.A. Beck,
Phys. Rev. C \textbf{47}, 2043 (1993).}

\bibitem{12} \textsf{R.N. Sagaidak, G.N. Kniajeva and I.M.
Itkis, Phys. Rev. C \textbf{68}, 014603 (2003).}


\bibitem{14} \textsf{ D.E. DiGregorio, M. di Tada and, D.
Abriola, Phys. Rev. C \textbf{39}, 516 (1989); and references
therein.}

\bibitem{16} \textsf{ J.O Fernandez Niello, M. di Tada and
A.O. Macchiavelli, Phys. Rev. C \textbf{43}, 2303 (1991).}

\bibitem{17} \textsf{ E. Martinez-Quiroz, E.F. Aguilera and
J.J. Kolata, Phys. Rev. C \textbf{63}, 054611 (2001).}

\bibitem{18} \textsf{ A. Mukherjee, M. Dasgupta and D.J.
Hinde, Phys. Rev. C \textbf{66} (2002) 34607.}

\bibitem{19} \textsf{L. Corradi, S.J. Skorka and U. Lenz,
Zeitschrift fur Physik, A \textbf{335}, 55 (1990).}


\bibitem{21} \textsf{D. J. Hinde, A. C. Berriman, R. D.Butt,
et al., J. Nucl. Rad. Sci, \textbf{3} No.1, 31 (2002); Eur. Phys.
J. A \textbf{13}, 149 (2002).}

\bibitem{22} \textsf{ Huanqiao Zhang, Jincheng Xu and Zuhua
Liu, Phys. Rev. C \textbf{42}, 1086 (1990).}

\bibitem{22a} \textsf{ B. B. Back, R. R. Betts, et al., Phys. Rev. C \textbf{32}, 195 (1985).}

\bibitem{23} \textsf{A.M. Stefanini, D. Ackermann and L.
Corradi, Phys. Rev. C \textbf{52}, 1727 (1995).}

\bibitem{24} \textsf{D.J. Hinde, A.C. Berriman and M.
Dasgupta, Phys. Rev. C \textbf{60}, 054602 (1999).}

\bibitem{25} \textsf{H.Timmers, D.Ackermann, S.Beghini, L.Corradi,
J.H.He, G.Montagnoli, F.Scarlassara, A.M.Stefanini, N.Rowley,
Nucl. Phys. \textbf{A633} 421 (1998). }


\bibitem{26} \textsf{P.H. Stelson, H.J. Kim and M. Beckerman,
Phys. Rev. C \textbf{41}, 1584 (1990).}

\bibitem{27} \textsf{P.R.S. Gomes, I.C. Charret and R. Wanis,
Phys. Rev. C \textbf{49}, 245 (1994).}

\bibitem{28} \textsf{ C.R. Morton, A.C. Berriman and R.D.
Butt, Phys. Rev. C \textbf{62}, 24607 (2000).}

\bibitem{29} \textsf{ R.D. Butt, D.J. Hinde, M. Dasgupta,
Phys. Rev. C \textbf{66}, 44601 (2002).}

\bibitem{31} \textsf{K. Nishio, H. Ikezoe, S. Mitsuoka, Phys.
Rev. C \textbf{62}, 014602 (2000).}

\bibitem{32} \textsf{ K.E. Zyromski, W. Loveland, G.A.
Souliotis, Phys. Rev. C \textbf{63}, 024615 (2001).}

\bibitem{33} \textsf{S.Mitsuoka, H. Ikezoe, K. Nishio and J. Lu, Phys. Rev. C \textbf{62}, 054603
(2000).}

\bibitem{34} \textsf{J.F.Liang, D.Shapira, C.J.Geene, et.
al.,Phys. Rev. Lett. \textbf{91}, 152701 (2003), and references
therein. }

\bibitem{35} \textsf{A.J. Pacheco, J.O Fernandez Niello and
D.E DiGregorio, Phys. Rev. C \textbf{45}, 2861 (1992).}


\bibitem{36} \textsf{A.M. Stefanini, G. Fortuna, R. Pengo et al., Nucl. Phys. A
\textbf{456}, 509 (1986).}

\bibitem{37} \textsf{A.M. Stefanini, L. Corradi, A.M. Vinodkumar et al.,
Phys. Rev. C \textbf{62}, 14601 (2000).}

\bibitem{Wang04} \textsf{Ning Wang, Zhuxia Li, Xizhen Wu, et al., Phys. Rev. C \textbf{69}, 034608 (2004).}


\end{thebibliography}
\end{document}